\documentclass[11pt,a4paper]{article}

\usepackage[T1]{fontenc}
\usepackage[utf8]{inputenc}
\usepackage{lmodern}
\usepackage{amsmath,amssymb,mathtools}
\usepackage[margin=1in]{geometry}
\usepackage{booktabs}
\usepackage{graphicx}
\usepackage{microtype}
\usepackage{xcolor}
\usepackage[colorlinks=true,linkcolor=magenta,citecolor=magenta,urlcolor=magenta,breaklinks=true]{hyperref}

\usepackage[nameinlink,noabbrev]{cleveref}

\newcommand{\PT}{\mathcal{PT}}
\newcommand{\dd}{\,\mathrm{d}}

\newcommand{\ii}{\mathrm{i}}
\newcommand{\B}{\mathrm{B}}
\newcommand{\C}{\mathbb{C}}

\title{\texorpdfstring{$\PT$}{PT}-Symmetric GUP-Modified Single- and Double-Well Quartic Oscillators}
\author{S. Zarrinkamar\\[2pt]
\small Department of Basic Sciences, Ga. C., Islamic Azad University, Garmsar, Iran\\
\small \href{mailto:saber.zarrinkamar@iau.ac.ir, zarrinkamar.s@gmail.com}{saber.zarrinkamar@iau.ac.ir, zarinkamar.s@gmail.com}\\
\small ORCID: \href{https://orcid.org/0000-0001-9128-4624}{0000-0001-9128-4624}}
\date{}

\begin{document}
\maketitle

\begin{abstract}
We study the GUP-deformed quartic oscillator $H_\eta=p^2+\eta p^4+x^4$ and distinguish its real-axis realization from complex-contour realizations. A contour chosen to remain in the decay sectors compatible with the undeformed quartic oscillator gives low-lying eigenvalues that approach the ordinary quartic spectrum as $\eta\to0^+$. This limiting-compatible contour does not, by itself, produce a low-level $\PT$-symmetry-breaking transition. To investigate exceptional points while retaining the same quartic leading behavior, we then introduce the explicitly $\PT$-symmetric extension
$H_{\eta,a,g}=p^2+\eta p^4+x^4-a x^2+\ii g x^3$ and fix $a=4$, for which the real part of the potential has a double-well form. Harmonic-oscillator-basis diagonalization with basis dimensions up to $N=100$ yields a converged exceptional-point branch: the critical coupling increases from $g_{\mathrm{EP}}\simeq0.10222364$ at $\eta=0$ to $g_{\mathrm{EP}}\simeq0.15603309$ at $\eta=0.1$. Across the transition, the two lowest real eigenvalues coalesce and continue as a complex-conjugate pair. The exceptional-point results belong to this stated double-well extension, not to the original single-well Hamiltonian alone.
\end{abstract}


\section{Introduction}

Although non-Hermitian Hamiltonians have long appeared in scattering theory and in effective descriptions of open systems, the modern study of parity-time ($\PT$) symmetry was transformed by the observation that non-Hermitian Hamiltonians can possess real spectra when their symmetry and boundary conditions are compatible. The seminal work of Bender and Boettcher established a broad class of such problems, and subsequent studies developed the roles of pseudo-Hermiticity, antilinear symmetries, supersymmetric constructions, and Lie-algebraic methods \cite{1998 Junker,Bender1998,1999 Znojil,2000 JPA Bagchi and Roy,2000 PLA Bagchi and Quesne,2001 Znojil,2001 Bagchi Znojil contin,2002 Bagchi quen Lie,2002 Fityo,2002 Mostafazadeh Pseudo-supersymmetric,2002 Mostafazadeh Pseudo-Hermiticity vs. PT,2002 Mostafazadeh Pseudo-Hermiticity vs. PT 2,2003 Mostafazadeh Pseudo-Hermiticity CPT,2003 Mostafazadeh Pseudo-Hermiticity,2006 Sinha and roy,2010 Bagchi rational extend,2015 Ahmed pla,2017 Levai,2022 shape invariance scarf EPL,Bender2024}. A central lesson of this literature is that formal invariance under $\PT$ does not by itself guarantee a real spectrum: the operator domain and the boundary conditions are essential parts of the spectral problem.\\
The interest in $\PT$ symmetry has also extended beyond formal quantum-mechanical models. Optical, photonic, electronic, and wave systems have provided experimental settings in which gain-loss balance, exceptional-point behavior, and related non-Hermitian effects can be investigated \cite{Ruter,Konotop,NatureChang,NaturePeng,NatureAshida,NatureAssawaworrarit,NatureWeimann,NatChen,NatureFan,NatureCao,NatureLiu,NatureFritzsche,anti 2025}. These developments make it useful to study carefully defined model problems whose spectral boundary conditions can be controlled analytically. They do not, however, remove the need to distinguish a formally $\PT$-invariant differential expression from a particular non-Hermitian realization of that expression. \\
Independently, generalized uncertainty principles (GUPs) and minimal-length models have been studied in several approaches to quantum gravity, string-inspired physics, black-hole physics, and doubly special relativity \cite{Konishi,Maggiore,Scardigli, Kempf1995, Bosso2017, Bosso2018, Vagenas PRD 2004,Cortes,Kempf,Hossenfelder,Ali,Vagenas 2009,Review 2015,Vagenas PRL 2008,Casadio 2015,Lambiase,Vagenas 2010 Relativistic,Nozari Pedram}. A commonly used first-order representation introduces a canonical momentum $p$ and a deformed momentum
\begin{equation}
 P=p\left(1+\frac{\beta}{3}p^2\right)+O(\beta^2),
 \label{eq:intro_gup_rep}
\end{equation}
which generates a positive $p^4$ correction in the kinetic energy. In one spatial dimension this correction changes the stationary Schr\"odinger equation from second to fourth order. Related links between minimal-length deformations and non-Hermitian or $\PT$-symmetric quantum mechanics have already been discussed, notably by Bagchi and Fring \cite{BagchiFring2009,BagchiFringComment2009}. Dey, Fring, and Gouba further studied $\PT$-symmetric non-commutative spaces whose uncertainty relations imply minimal lengths, areas, or volumes \cite{DeyFringGouba2012}. These works establish relevant background, although their specific models differ from the quartic oscillator considered here. The general statement that GUP and $\PT$-symmetric ideas can be combined is therefore not the novelty claimed here; the question is what follows for a specific fourth-order spectral problem with explicitly stated complex-contour boundary conditions.\\
The quartic oscillator is a natural test case. Its conventional spectral properties have been investigated using perturbation theory, nonperturbative methods, unitary transformations, and the analysis of Stokes multipliers and quantization conditions \cite{2004 JPA unitary transformation,2006 JMP,2006 PRE,2007 PRA diffuse dynamics,2008 PRE entanglement,2011 JPA Osherov,2014 PRD Liouville equation,2016 PLA qubit,2020 PRa,2024 Path Integral,2024 JHEP WKB}. The quartic potential is simple enough to permit a transparent asymptotic analysis while remaining nontrivial. In the present problem, the positive $p^4$ term has a further consequence: for every fixed nonzero deformation parameter it controls the leading large-$|x|$ asymptotics, even when its effect on low-energy states is perturbatively small. More broadly, higher-derivative Hamiltonians have been studied in other settings, including integrable systems and higher charges of affine Toda models \cite{FringTurner2023Scattering,FringTurner2023Ghosts}; these works provide context for higher-order kinetic operators but do not derive the spectrum of the present GUP oscillator.\\
Here we formulate the GUP-deformed quartic oscillator in dimensionless form and define a specific $\PT$-invariant complex-contour problem. We derive the four leading asymptotic exponential branches, determine which two-dimensional decaying subspaces are selected at the two ends of the contour, and state the spectral consequences that follow from the antilinear symmetry. We also derive a first-order low-momentum WKB correction and formulate a matching determinant for computing the spectrum. The distinction between exact analytical consequences and numerical questions is maintained throughout: no numerical eigenvalues are presented unless they have passed convergence and independent-method checks. In particular, the present analysis supplies criteria for diagnosing a real spectrum and for identifying possible $\PT$-symmetry breaking, but does not assert that a transition has been observed.\\
The paper is organized as follows. Section~2 introduces the dimensionless GUP-deformed quartic oscillator and analyzes its leading asymptotic branches, decay sectors, and the consequences of $\PT$ symmetry. Section~3 derives the low-momentum WKB correction and discusses the real-axis benchmark and complex-contour cross-check. Section~4 formulates the matching determinant for the specified contour and examines what can, and cannot, be concluded about the spectrum from this construction. Section~5 considers a separate $\PT$-symmetric double-well extension, where exceptional points can be investigated as a function of the deformation and coupling parameters. Section~6 summarizes the results, discusses their limitations, and outlines directions for further work.
\section{GUP-deformed quartic oscillator}
We adopt the first-order momentum representation \cite{Maggiore,Scardigli,Vagenas PRD 2004,Kempf,Hossenfelder,Ali,Review 2015}.
\begin{equation}
 P=p\left(1+\frac{\beta}{3}p^2\right)+O(\beta^2),
 \qquad p=-\ii\hbar\frac{\dd}{\dd X}.
 \label{eq:GUP}
\end{equation}
To first order in $\beta$,
\begin{equation}
 \frac{P^2}{2m}=\frac{p^2}{2m}+\frac{\beta}{3m}p^4+O(\beta^2).
 \label{eq:kinetic}
\end{equation}
For $V(X)=\lambda X^4$, with $\lambda>0$, the stationary equation is
\begin{equation}
 \left[\frac{\beta\hbar^4}{3m}\frac{\dd^4}{\dd X^4}
 -\frac{\hbar^2}{2m}\frac{\dd^2}{\dd X^2}+\lambda X^4\right]\Psi(X)
 =\mathcal E\Psi(X).
 \label{eq:dimensional}
\end{equation}
The parameter $\beta$ has dimensions of inverse momentum squared. Since the momentum representation in \cref{eq:GUP} is truncated at first order, the fourth-order equation should be understood as the associated effective spectral model; its far-field behavior at arbitrarily large momentum need not represent a controlled truncation of a more complete GUP theory.
Introducing
\begin{equation}
 L=\left(\frac{\hbar^2}{2m\lambda}\right)^{1/6},\qquad x=\frac{X}{L},\qquad
 E=\frac{\mathcal E}{\hbar^2/(2mL^2)},\qquad
 \eta=\frac{2\beta\hbar^2}{3L^2}.
 \label{eq:scaling}
\end{equation}
Then the dimensionless equation is
\begin{equation}
 \eta\psi^{(4)}(x)-\psi''(x)+(x^4-E)\psi(x)=0,\quad \eta>0.
 \label{eq:main}
\end{equation}
Equivalently, $H_\eta=p^2+\eta p^4+x^4$ with $p=-\ii\dd/\dd x$. The numerical values of $\eta$ used below are dimensionless values in this scaled model, not estimates of a physically measured minimal-length parameter. In particular, the larger values (including $\eta=0.05$, $0.1$, $0.5$, and $1$ in the single-well benchmark, and values up to $0.1$ in the exceptional-point scan) are deliberately exaggerated relative to a perturbatively small GUP correction. They are included to make the spectral trends and the dependence of the exceptional-point threshold visible on the plots and in the tables. Quantitative predictions for a particular physical system would require fixing the dimensional scales and checking that the first-order GUP representation remains valid in the regime considered. \\
The differential expression in \cref{eq:main} is invariant under $\PT$, acting as $x\mapsto-x$, $\ii\mapsto-\ii$. On the real axis, however, the usual self-adjoint realization is Hermitian and has a real discrete spectrum. A non-Hermitian $\PT$-symmetric problem requires complex-contour boundary conditions; the contour and its endpoint decay subspaces are part of the definition of the operator.
\subsection{Leading asymptotics and decay sectors}
For fixed $\eta>0$ and large $|x|$, the leading equation is
\begin{equation}
 \eta\psi^{(4)}(x)+x^4\psi(x)\simeq0.
 \label{eq:leading}
\end{equation}
With $\psi\sim\exp S(x)$, the leading WKB balance gives
\begin{equation}
 \eta[S'(x)]^4+x^4\simeq0,\qquad
 S'_k(x)=\eta^{-1/4}x\exp\left[\frac{(2k+1)\pi\ii}{4}\right],\quad k=0,1,2,3.
\end{equation}
Integration yields
\begin{equation}
 \psi_k(x)\sim\exp\left[\frac{x^2}{2\eta^{1/4}}
 \exp\left(\frac{(2k+1)\pi\ii}{4}\right)\right],\quad k=0,1,2,3.
 \label{eq:asymptotic}
\end{equation}
For $x=re^{\ii\theta}$, the leading modulus is
\begin{equation}
 |\psi_k|\sim\exp\left[\frac{r^2}{2\eta^{1/4}}
 \cos\left(2\theta+\frac{(2k+1)\pi}{4}\right)\right],
 \label{eq:modulus}
\end{equation}
so branch $k$ decays in angular intervals satisfying
\begin{equation}
 \cos\left(2\theta+\frac{(2k+1)\pi}{4}\right)<0.
 \label{eq:decay}
\end{equation}
The centers of these intervals are
\begin{equation}
 \theta_{k,m}=\frac{\pi}{2}-\frac{(2k+1)\pi}{8}+m\pi,\qquad m\in\mathbb Z,
 \label{eq:centers}
\end{equation}
and each interval has angular width $\pi/2$. Representatives are $67.5^\circ$, $22.5^\circ$, $-22.5^\circ$, and $-67.5^\circ$, with the opposite directions obtained by adding $180^\circ$. The $\PT$ transformation maps a ray at angle $\theta$ to one at $\pi-\theta$ modulo $2\pi$. A $\PT$-symmetric boundary-value problem must pair the endpoint conditions under this mapping. Since \cref{eq:main} is fourth order, a standard discrete problem requires a two-dimensional decaying subspace at each endpoint. The sign condition in \cref{eq:decay} classifies individual exponential branches; it does not by itself prove that every apparent pair of sectors defines an admissible spectral problem. The subspaces must be continued consistently and checked numerically.\\
For a contour that preserves a controlled $\eta\to0^+$ limit, we choose asymptotic endpoint rays $\theta_R=15^\circ$ and $\theta_L=165^\circ$. They satisfy $\theta_L=\pi-\theta_R$ and are therefore paired by $x\mapsto-x^*$ under $\PT$. The contour used numerically is
\begin{equation}
 x(t)=t+\ii\tan(15^\circ)\frac{t^2}{\sqrt{1+t^2}},\qquad -\infty<t<\infty,
 \label{eq:contour}
\end{equation}
which passes through the origin and is invariant under $\PT$. On the right ray the decaying branches are $k=1,2$, and on the left ray they are also $k=1,2$. These are the same branches selected on the real axis, and the chosen contour can be deformed to the real axis without crossing a Stokes boundary. This choice is deliberate: the earlier $50^\circ/130^\circ$ pair selects different fourth-order sectors and does not furnish the same controlled connection to the ordinary quartic-oscillator boundary conditions. The contour in \cref{eq:contour} is therefore appropriate for the limiting-spectrum question, though it is not expected to generate a distinct $\PT$-broken phase by contour rotation alone.
\subsection{Spectral consequences of $\PT$ symmetry}
Let $\mathcal{C}$ denote the specified contour and let the operator domain include the two decaying conditions at each end. If this domain is invariant under $\PT$, then from $H_\eta\psi=E\psi$ it follows that $H_\eta(\PT\psi)=E^*(\PT\psi)$. Consequently, an eigenvalue is either real or belongs to a complex-conjugate pair $E,E^*$. This implication does not force every eigenvalue to be real. A real eigenvalue with a $\PT$-invariant eigenfunction is conventionally associated with the unbroken phase; a non-real conjugate pair signals broken $\PT$ symmetry for the corresponding states. The symmetry statement is conditional on the contour and operator domain being invariant under $\PT$.\\
For a parameter-dependent spectral problem, an exceptional point can occur when two or more eigenvalues and their eigenvectors coalesce. Numerically, a candidate transition must be supported by simultaneous evidence of eigenvalue coalescence, loss of reality beyond the candidate point, and stable results under refinement. A small imaginary part at finite precision is not sufficient evidence of symmetry breaking, just as an apparently real eigenvalue at one resolution is not proof of an unbroken phase. Since the present model has only the GUP parameter $\eta$ after scaling, a scan in $\eta$ is meaningful only after the selected contour and its endpoint subspaces have been fixed and the determinant has been shown to converge.\\
The limit $\eta\to0^+$ is singular. At every fixed nonzero $\eta$, the fourth derivative controls the leading far-field equation, whereas setting $\eta=0$ reduces the differential order to two. The undeformed quartic problem therefore cannot be recovered uniformly in $x$ from \cref{eq:asymptotic}.
\section{Low-momentum WKB correction}
For the perturbative low-momentum branch, let $Q(x)=E-x^4$. In the classically allowed interval $Q\geq0$, the local momentum $q$ satisfies
\begin{equation}
 q^2+\eta q^4=Q,\qquad q=Q^{1/2}-\frac{\eta}{2}Q^{3/2}+O(\eta^2).
 \label{eq:q}
\end{equation}
The undeformed leading action is
\begin{equation}
 I_0(E)=\int_{-E^{1/4}}^{E^{1/4}}\sqrt{E-x^4}\,\dd x
 =2E^{3/4}J_4,\qquad
 J_4=\int_0^1\sqrt{1-t^4}\,\dd t=\frac14\B\!\left(\frac14,\frac32\right),
 \label{eq:I0}
\end{equation}
with $I_0(E_n^{(0)})=\pi(n+1/2)$. The first-order change in the action is $-\eta E^{7/4}K_4$, where
\begin{equation}
 K_4=\int_0^1(1-t^4)^{3/2}\,\dd t=\frac14\B\!\left(\frac14,\frac52\right),\qquad \frac{K_4}{J_4}=\frac67.
\end{equation}
Expanding the Bohr--Sommerfeld condition to first order gives
\begin{equation}
 E_n=E_n^{(0)}+\frac47\eta\bigl(E_n^{(0)}\bigr)^2+O(\eta^2).
 \label{eq:wkb}
\end{equation}
Here $E_n^{(0)}$ denotes the undeformed WKB energy, not the exact quartic-oscillator eigenvalue. Also, The function \(B(a,b)\) denotes the Euler beta function, defined by
\begin{equation}
B(a,b)
=
\int_0^1 t^{a-1}(1-t)^{b-1}\,dt
=
\frac{\Gamma(a)\Gamma(b)}{\Gamma(a+b)},
\qquad
\operatorname{Re}(a)>0,\quad
\operatorname{Re}(b)>0,
\label{eq:beta-function}
\end{equation}
where \(\Gamma(z)\) is the Euler gamma function. This result is a semiclassical estimate for the low-momentum branch; it does not determine the complex-contour $\PT$ spectrum.
\subsection{Real-axis benchmark and the small-\texorpdfstring{$\eta$}{eta} limit}
The complex-contour problem above and the standard real-axis problem must not be identified with one another. To obtain a well-defined comparison with the undeformed quartic oscillator, we also solve the self-adjoint realization of
\begin{equation}
 H_\eta\psi=E\psi,\qquad
 H_\eta=-\frac{\dd^2}{\dd x^2}+\eta\frac{\dd^4}{\dd x^4}+x^4,
 \qquad \psi\in L^2(\mathbb R),\quad \eta\geq0.
 \label{eq:realaxis}
\end{equation}
For $\eta>0$, the Fourier-space kinetic energy is $k^2+\eta k^4\geq0$, so the real-axis operator is bounded below and self-adjoint on its standard domain. Its eigenvalues are therefore real. This benchmark can test the $\eta\to0^+$ limit, but it cannot exhibit a real-to-complex exceptional-point transition: such a transition is excluded for this self-adjoint realization.

We computed the low-lying real-axis spectrum using Chebyshev collocation on $[-L,L]$. For $\eta>0$, the fourth-order equation was approximated with decaying-endpoint conditions $\psi(\pm L)=\psi'(\pm L)=0$; the $\eta=0$ reference was calculated with the corresponding second-order Dirichlet problem. The reported values were checked by increasing the collocation order and changing the truncation interval. For example, the first four eigenvalues at $\eta=0.1$ were unchanged to the displayed digits when the discretization was increased from $N=60$ to $N=120$ and $L$ varied from $7$ to $9$. The values below are dimensionless.
\begin{table}[htbp]
\centering
\caption{Lowest real-axis eigenvalues of $H_\eta=p^2+\eta p^4+x^4$. The $\eta=0$ row is the ordinary quartic oscillator. The larger $\eta$ values are illustrative, deliberately enhanced to make the trend visible; they should not be read as phenomenological estimates.}
\label{tab:realaxis}
\begin{tabular}{c r r r r}
\toprule
$\eta$ & $E_0$ & $E_1$ & $E_2$ & $E_3$\\
\midrule
$0$      & 1.06036209 & 3.79967303 & 7.45569794 & 11.64474551\\
$10^{-4}$& 1.06050389 & 3.80069094 & 7.45913953 & 11.65282261\\
$10^{-3}$& 1.06177810 & 3.80982632 & 7.48997501 & 11.72504489\\
$10^{-2}$& 1.07432933 & 3.89878807 & 7.78587765 & 12.40650325\\
$0.05$   & 1.12665980 & 4.25390836 & 8.91224375 & 14.87990790\\
$0.1$    & 1.18614511 & 4.63649783 & 10.06799725 & 17.31349073\\
$0.2$    & 1.29182686 & 5.28356402 & 11.95143381 & 21.17221720\\
$0.5$    & 1.54930720 & 6.77188528 & 16.12250465 & 29.50890992\\
$1$      & 1.88191791 & 8.61024310 & 21.13806573 & 39.37531742\\
\bottomrule
\end{tabular}
\end{table}
\begin{figure}[htbp]
\centering
\includegraphics[width=0.82\textwidth]{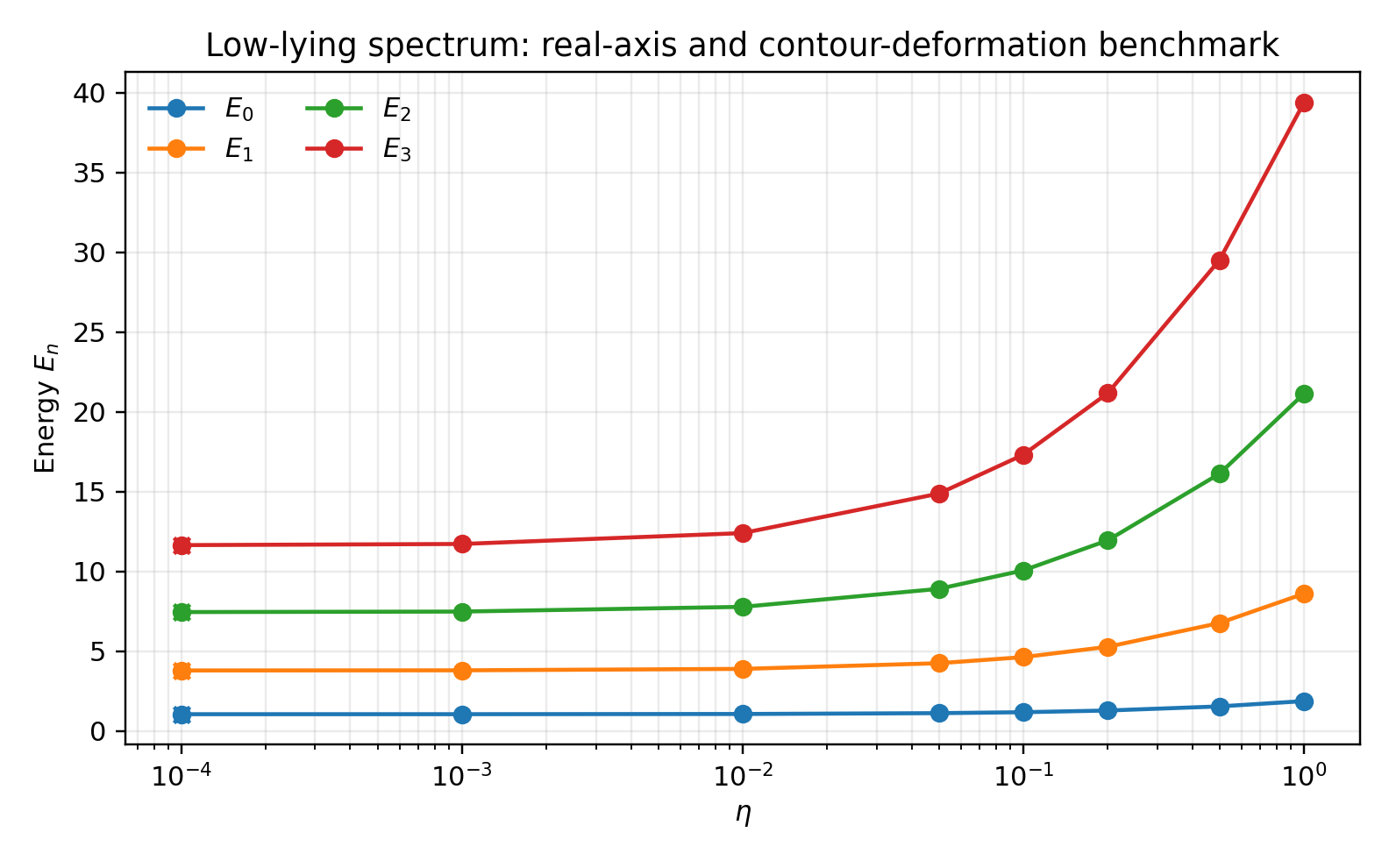}
\caption{Lowest four eigenvalues as functions of $\eta$. The real-axis values and the $15^\circ/165^\circ$ contour-deformation values agree within numerical accuracy. The crosses indicate the undeformed quartic-oscillator levels at $\eta=0$.}
\label{fig:realaxis}
\end{figure}
The numerical trend is consistent with convergence of the low-lying real-axis levels to those of the undeformed quartic oscillator as $\eta$ decreases. In particular, the first four eigenvalues at $\eta=10^{-4}$ differ from the $\eta=0$ values by approximately $1.42\times10^{-4}$, $1.02\times10^{-3}$, $3.44\times10^{-3}$, and $8.08\times10^{-3}$, respectively. This provides a numerical benchmark for the low-energy branch. It does not prove uniform convergence of the eigenfunctions in the far-field region, where the order of the differential equation changes at $\eta=0$.
\subsection{Complex-contour cross-check}
We independently discretized the differential operator on the contour in \cref{eq:contour} by Chebyshev collocation in the real parameter $t$. The derivatives were transformed using $\dd/\dd x=(x'(t))^{-1}\dd/\dd t$, and the endpoint rows were replaced by the asymptotic two-dimensional decaying-subspace conditions derived from the WKB branches. For $\eta=10^{-4},10^{-3},10^{-2},0.1$, the calculation used $N=100$ Chebyshev intervals and $L=8$; for $\eta=0.5,1$, it used $N=120$ and $L=10$. The low-lying eigenvalues agree with the real-axis benchmark within the numerical accuracy. The maximum absolute imaginary part among the first four computed eigenvalues is below $10^{-8}$ for all six sampled values of $\eta$. This is a numerical cross-check of contour-deformation equivalence, not evidence for a separate non-Hermitian phase. To avoid duplicating the real-axis values, we report the contour comparison graphically in \cref{fig:realaxis} rather than reproducing the same spectrum in a second table.
\section{Matching determinant for a specified contour}
For a selected contour and a fixed $\eta>0$, we rewrite \cref{eq:main} as a first-order system for
\begin{equation}
 Y(x)=\bigl(\psi,\psi',\psi'',\psi'''\bigr)^{\mathsf T}.
\end{equation}
Let $Y_L^{(1,2)}$ be a basis satisfying the left-end decay condition and $Y_R^{(1,2)}$ a basis satisfying the right-end decay condition. Analytically continue all four solutions to a common matching point, taken as $x=0$ for the contour in \cref{eq:contour}, and define
\begin{equation}
 D(E;\eta)=\det\left[Y_L^{(1)},Y_L^{(2)},Y_R^{(1)},Y_R^{(2)}\right]_{x=0}.
 \label{eq:det}
\end{equation}
Zeros of $D$ are candidate eigenvalues for that particular boundary-value problem. Their reliability depends on accurate initialization of the decaying subspaces. At a finite endpoint $x_R$, the leading asymptotic logarithmic derivatives are obtained from $S'_k(x_R)=\eta^{-1/4}x_R\exp[(2k+1)\pi\ii/4]$; higher-order asymptotic terms should be included before a production calculation. For the revised $15^\circ/165^\circ$ contour, the right- and left-end subspaces are both spanned by the decaying modes $k=1,2$. At a finite endpoint, we impose the two-dimensional asymptotic subspace through linear boundary conditions constructed from the WKB derivative vectors of the selected branches. The numerical collocation implementation uses these asymptotic subspace conditions rather than simply setting $\psi$ and $\psi'$ to zero at each endpoint.\\
A reproducible calculation should (i) increase the endpoint radius, (ii) refine the integration or collocation resolution, (iii) increase the asymptotic initialization order, (iv) vary tolerances and arithmetic precision, and (v) compare the determinant zeros with an independent discretization. For every reported eigenvalue, the manuscript should give the changes under each refinement and the residual in the matching conditions. To test $\PT$ symmetry, the computation should also check that non-real roots appear in conjugate pairs within numerical uncertainty. A claim of an exceptional point requires tracking the same eigenvalue branches as $\eta$ varies and demonstrating convergence of their coalescence, not merely observing a close pair at one parameter value.
\subsection{Exceptional-points}
The revised contour gives a well-defined comparison with the undeformed quartic oscillator. The asymptotic rays at $15^\circ$ and $165^\circ$ lie in decay sectors that are compatible with the real-axis problem at both $\eta>0$ and $\eta=0$. The numerical contour spectrum agrees with the real-axis spectrum for the first four levels over the scan $10^{-4}\leq\eta\leq1$, and the low-lying levels remain real to numerical precision. The small-$\eta$ values approach the standard quartic levels, as displayed in Table~\ref{tab:realaxis} and Fig.~\ref{fig:realaxis}.\\
No exceptional point is found in this scan for the selected contour. This is not a claim that all possible $\PT$-symmetric realizations of the fourth-order equation have real spectra. Rather, the contour has deliberately been chosen within the same Stokes sectors as the Hermitian realization; its spectral problem is a contour deformation of that realization. Varying $\eta$ alone in this setup does not generate a distinct non-Hermitian phase in the low-lying spectrum.
The earlier $50^\circ/130^\circ$ contour selects different decaying fourth-order branches. Preliminary collocation tests for that sector-switched problem were sensitive to endpoint radius and resolution, and did not produce stable eigenvalue branches. We therefore do not report its apparent complex roots as physical data. Moreover, that sector choice cannot simply be assumed to connect to the ordinary quartic oscillator when $\eta\to0^+$, because the fourth-order Stokes sectors and the second-order decay sectors do not match uniformly.\\
The three original questions can now be answered with the necessary distinction. First, the lowest eigenvalues remain real over the tested interval for the limiting-compatible contour. Second, no real branches meet to form complex-conjugate pairs in that interval; an exceptional point has not been observed. Third, the low-lying spectrum approaches the undeformed quartic spectrum as $\eta\to0^+$, although this numerical spectral convergence does not establish uniform convergence of eigenfunctions in the far field. To investigate a genuine $\PT$-breaking transition while retaining a controlled quartic limit, one must introduce an additional non-Hermitian deformation or a second physical parameter and then verify its spectrum independently. It would be misleading to infer such a transition merely by rotating the contour across a Stokes boundary, since that changes the boundary-value problem itself.
\section{A PT-symmetric double-well extension and exceptional points}
\label{sec:EPextension}
The single-well Hamiltonian $H_\eta=p^2+\eta p^4+x^4$ with the limiting-compatible boundary conditions above does not show a low-lying exceptional point in the parameter scan. Exceptional points and the possible loss of spectral reality are established features of other complex quartic spectral problems, but they depend on the precise operator and boundary conditions \cite{DelabaerePham1998}. To test whether the GUP kinetic correction shifts a genuine $\PT$-breaking transition, we therefore study a separate, explicitly defined extension,
\begin{equation}
 H_{\eta,a,g}=p^2+\eta p^4+x^4-a x^2+\ii g x^3,
 \qquad p=-\ii\frac{\dd}{\dd x},\quad \eta\geq0,\quad a,g\in\mathbb R.
 \label{eq:EP-Hamiltonian}
\end{equation}
The parameters $a$ and $g$ are not part of the original single-well model. We fix $a=4$ throughout the numerical study. The real part $x^4-4x^2$ has two minima and a central barrier, while the imaginary cubic term is $\PT$-symmetric: parity changes the sign of $x^3$, and time reversal changes $\ii$ to $-\ii$, leaving $\ii g x^3$ invariant. The kinetic terms $p^2$ and $p^4$ are also invariant under $\PT$.
We represent the Hamiltonian in the harmonic-oscillator basis $\{|n\rangle\}_{n=0}^{N-1}$, with
\begin{equation}
 x=\frac{a_{\rm ho}+a_{\rm ho}^{\dagger}}{\sqrt{2}},\qquad
 p=\frac{\ii(a_{\rm ho}^{\dagger}-a_{\rm ho})}{\sqrt{2}},
 \qquad [a_{\rm ho},a_{\rm ho}^{\dagger}]=1.
 \label{eq:ho-basis}
\end{equation}
The finite matrix is the projection of the operator in \cref{eq:EP-Hamiltonian} onto the first $N$ oscillator states. Matrix elements of $x^k$ and $p^k$ are formed before projection, using an auxiliary basis enlarged by four states so that the powers through fourth order are not contaminated by an artificial truncation at the intermediate multiplication step. We diagonalize the resulting complex-symmetric matrix and follow the two lowest eigenvalue branches.

For a pair of eigenvalues $E_0(g)$ and $E_1(g)$ that are real below the transition, the exceptional point is located numerically by solving
\begin{equation}
 \operatorname{Re}\left[(E_0(g)-E_1(g))^2\right]=0.
 \label{eq:EP-condition}
\end{equation}
This condition is used together with branch tracking on either side of the candidate point: below it, the pair is real and split; above it, the pair has equal real parts and opposite imaginary parts. We also checked the normalized overlap of the two right eigenvectors. For $\eta=0.01$, the overlap tends to unity as the parameter approaches the candidate point from either side, consistent with eigenvector coalescence at an exceptional point. The numerical estimates below were stable as the basis dimension increased from $N=40$ to $N=100$.

\begin{table}[htbp]
\centering
\caption{Exceptional-point estimates for the $\PT$-symmetric double-well extension \cref{eq:EP-Hamiltonian} at $a=4$. Here $g_{\rm EP}$ is the critical imaginary-cubic coupling and $E_{\rm EP}$ is the coalescence energy. Values are rounded to six significant digits; the last column gives the change in $g_{\rm EP}$ between basis dimensions $N=40$ and $N=100$. Some dimensionless $\eta$ values are deliberately enhanced to make the trend visible and should not be interpreted as realistic GUP estimates.}
\label{tab:EPvalues}
\begin{tabular}{c c c c}
\toprule
$\eta$ & $g_{\rm EP}$ & $E_{\rm EP}$ & $|g_{\rm EP}^{(100)}-g_{\rm EP}^{(40)}|$\\
\midrule
$0$       & 0.102224 & $-1.46704$ & $2.5\times10^{-9}$\\
$0.001$   & 0.102779 & $-1.46268$ & $1.2\times10^{-9}$\\
$0.01$    & 0.107685 & $-1.42539$ & $1.3\times10^{-9}$\\
$0.05$    & 0.128897 & $-1.28637$ & $7.6\times10^{-12}$\\
$0.1$     & 0.156033 & $-1.14523$ & $2.2\times10^{-13}$\\
\bottomrule
\end{tabular}
\end{table}

\begin{figure}[htbp]
\centering
\includegraphics[width=0.98\textwidth]{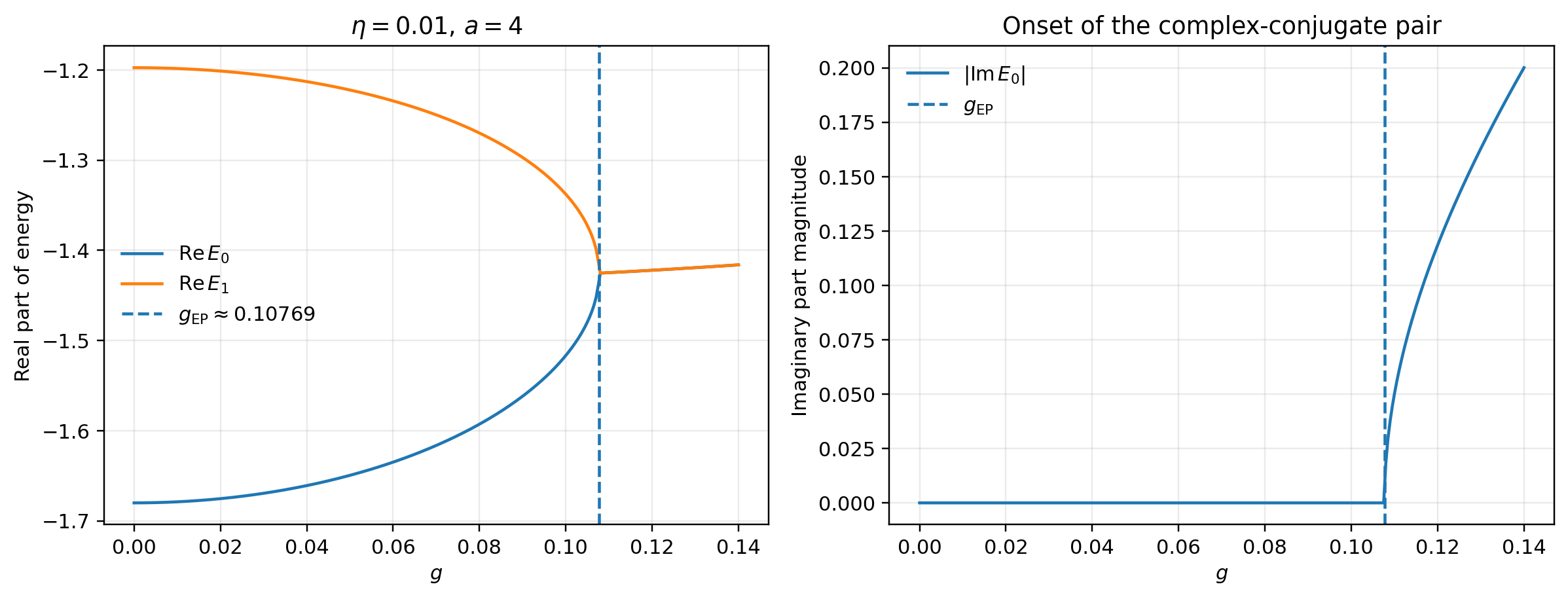}
\caption{Exceptional-point behavior for the double-well extension with $a=4$ and $\eta=0.01$. The two lowest real branches approach and coalesce at $g_{\rm EP}\simeq0.10769$; beyond the transition their real parts coincide and the imaginary parts form a complex-conjugate pair.}
\label{fig:EPscan}
\end{figure}

\begin{figure}[htbp]
\centering
\includegraphics[width=0.68\textwidth]{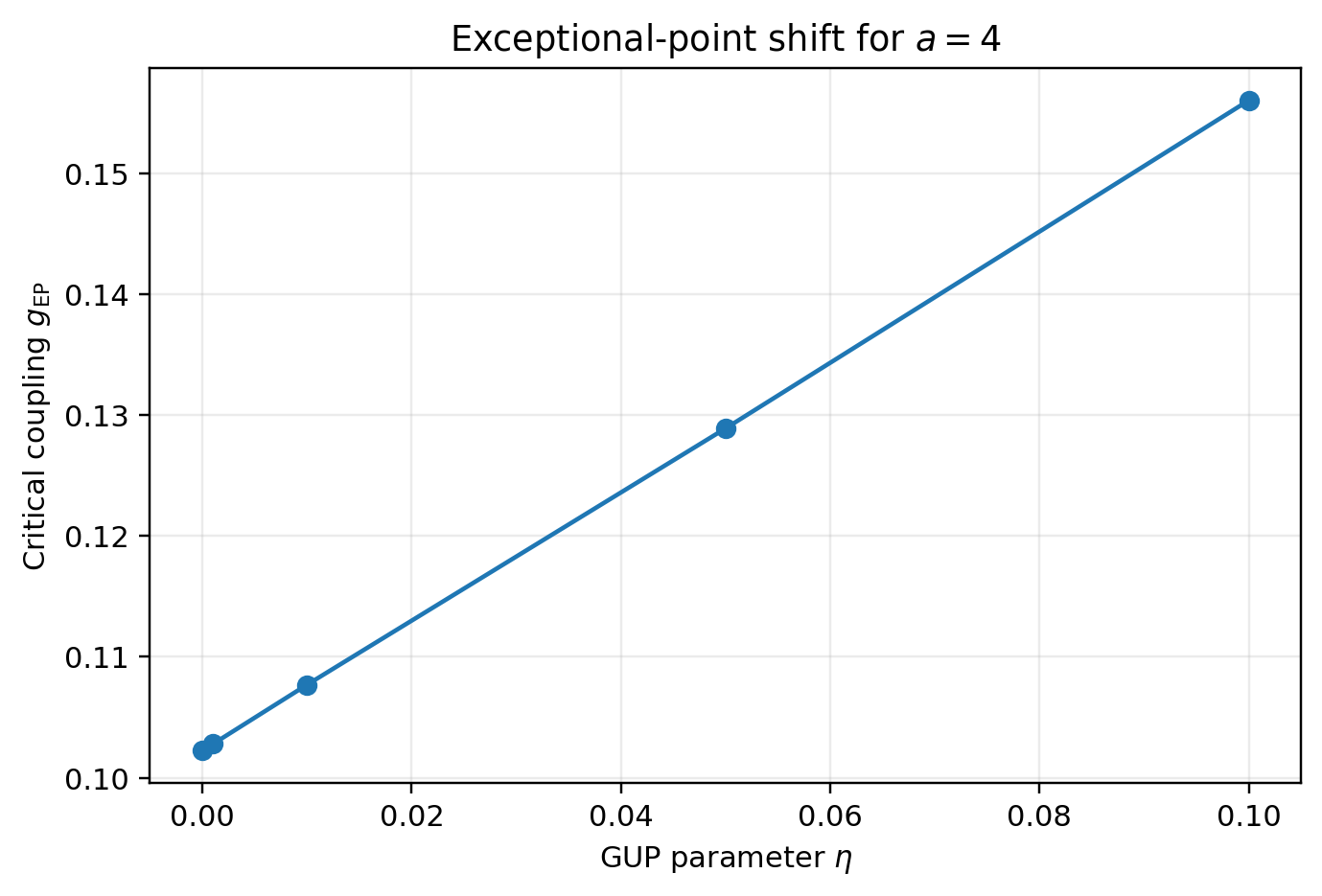}
\caption{Critical coupling $g_{\rm EP}$ versus the GUP parameter $\eta$ for the double-well extension at fixed $a=4$. The critical coupling increases over the sampled interval, indicating that the positive $p^4$ correction shifts the observed low-level transition to larger $g$.}
\label{fig:EPvsEta}
\end{figure}

For the representative value $\eta=0.01$, the first two levels are real at $g=0.102685$, just below the estimated critical value, with energies approximately $-1.499209$ and $-1.353999$. At $g=0.112685$, just above the transition, they form a complex-conjugate pair,
\begin{equation}
 E_{\pm}\simeq -1.424119\pm0.074295\,\ii.
 \label{eq:EP-pair-example}
\end{equation}
The coalescence energy and threshold vary with $\eta$, as shown in \cref{tab:EPvalues,fig:EPvsEta}. These calculations establish an exceptional-point transition for the specified extension and parameter slice. They do not establish such a transition for the original single-well Hamiltonian $p^2+\eta p^4+x^4$ without the additional $-a x^2+\ii g x^3$ terms.
\section{Discussion and conclusions}

We have separated two spectral questions that should not be conflated. For the original GUP-deformed single-well oscillator $H_\eta=p^2+\eta p^4+x^4$, the limiting-compatible contour yields a low-lying spectrum that agrees with the real-axis benchmark and approaches the ordinary quartic spectrum as $\eta\to0^+$. No low-lying exceptional point was found in the sampled interval for that realization. This is consistent with the fact that the chosen contour preserves the relevant decay sectors rather than defining a different $\PT$-broken phase.\\
To investigate a genuine transition, we introduced the distinct $\PT$-symmetric extension $H_{\eta,a,g}=p^2+\eta p^4+x^4-a x^2+\ii g x^3$ and fixed $a=4$. The harmonic-oscillator-basis calculations show a converged coalescence of the two lowest levels followed by a complex-conjugate pair. The estimated threshold rises from $g_{\rm EP}\simeq0.10222364$ at $\eta=0$ to $g_{\rm EP}\simeq0.15603309$ at $\eta=0.1$. Thus, for this specified double-well extension, the positive $p^4$ correction shifts the observed low-level exceptional point to larger values of the imaginary-cubic coupling.\\
These exceptional-point results are numerical and model-specific. The sampled $\eta$ values, especially the larger values included to make the trend visually clear, are deliberately exaggerated dimensionless values; they demonstrate the behavior of the scaled model rather than establish a realistic experimental range for the GUP parameter. They should not be attributed to the original single-well GUP Hamiltonian by itself, nor should the added double-well and imaginary-cubic terms be treated as consequences of the GUP representation. The contribution of the study is instead to identify a controlled limiting problem for the original operator and to provide a separate, explicit $\PT$-symmetric extension in which the effects of the GUP parameter on an exceptional-point transition can be quantified. The general association between minimal-length deformations and $\PT$ symmetry has been discussed previously \cite{BagchiFring2009}; the numerical results here concern the stated boundary-value problem and the specified extension.\\
Several extensions of the present analysis merit investigation. First, the calculation could be generalized to other forms of the generalized uncertainty principle, including higher-order momentum deformations and nonpolynomial deformation functions. Comparing these realizations would help distinguish spectral features that are robust under changes in the deformation prescription from those that depend on a particular model. Second, it would be useful to examine other interactions, such as sextic or mixed anharmonic terms and, where mathematically appropriate, Coulomb-type potentials, both with and without explicit $\PT$-symmetric perturbations. Such studies could clarify how the interaction and the minimal-length deformation jointly affect spectral reality, level crossings, and exceptional points. Third, systematic scans of exceptional-point locations and their scaling with the deformation parameter could identify changes in spectral branches and level spacings that distinguish different effective models. These patterns may help constrain minimal-length models, although their interpretation as physical signatures would require independent justification and a clearly specified observable. Finally, semiclassical quantization, perturbation theory, and high-precision numerical calculations should be combined to test the range of validity of the present approximations and to check the robustness of the numerical trends.
\section*{Declaration of Generative AI and AI-Assisted Technologies}
Generative AI tools were used to assist in the preparation of this manuscript, including language refinement, \LaTeX{} formatting, and the organization and presentation of the numerical results. The author reviewed and revised the AI-assisted material and takes full responsibility for the scientific content, interpretation of the results, and final version of the manuscript.
\section*{Aknowledgements}
The authors wishes to thank P. Ghalandari for several useful comments on the numerical points. 

\end{document}